\documentclass[aps, prl, twocolumn, showpacs, nobibnotes, 10pt]{revtex4-1}

\newcommand{\D}[2]{\frac{\partial #1}{\partial #2}}
\newcommand{\DD}[2]{\frac{\partial^2 #1}{\partial #2^2}}
\newcommand{\dd}{~\mathrm{d}}

\catcode`,\active

\catcode`\,12

\usepackage{graphicx, amsmath, amssymb, bm, mathtools}

\newcommand{\be}{\begin{equation}}
\newcommand{\ee}{\end{equation}}
\newcommand{\BE}{\begin{eqnarray}}
\newcommand{\EE}{\end{eqnarray}}

\begin{document}

\title{Noise-Induced Bistable States and Their Mean Switching Time in Foraging Colonies}

\author{Tommaso Biancalani}
\author{Louise Dyson}
\author{Alan J.~McKane}

\affiliation{Theoretical Physics Division, School of Physics and Astronomy, 
University of Manchester, Manchester M13 9PL, United Kingdom}

\begin{abstract}
We investigate a type of bistability occurring in population systems where noise not only causes transitions between stable states, but also constructs the states themselves. We focus on the experimentally well-studied system of ants choosing between two food sources to illustrate the essential points, but the ideas are more general. The mean time for switching between the two bistable states of the system is calculated. This suggests a procedure for estimating, in a real system, the critical population size above which bistability ceases to occur. 
\end{abstract}


\maketitle

Bistable systems, as their name implies, are systems which may reside in one of two states. Typically, these states are extremely stable, with rare transitions only occurring through the effects of noise (intrinsic or extrinsic) or external perturbations. 


The standard theoretical approach used to investigate bistability is to begin by modeling the system deterministically though a set of differential or difference equations. In the deterministic system there can be no transitions between steady states without the addition of noise to move the system from one state to the other. The theoretical literature examining this effect is enormous, with very many variants of this basic scenario having been investigated in considerable detail~{\cite{Gardiner2009}. The majority of these theoretical studies fail to use the noise structure appropriate to the system under consideration, and reverse the logical sequence of model building: the deterministic equations together with the correct form of the noise should follow from a model constructed at the microscale (see for instance~\cite{McKane2004} or~\cite{gillespie2013perspective}).

A bottom-up approach such as this is required to understand unexpected and non-intuitive results such as those seen when a chemical system with a single stable fixed point is driven to bistability at low molecule numbers~\cite{Togashi2001}. This recently discovered mechanism for bistability, so far only investigated in the context of biochemical reactions, is a result of the non-linear nature of the intrinsic noise~\cite{Togashi2001, Ohkubo2007, Biancalani2012, Remondini2013, Popovic2013}. In this type of bistability, the noise is responsible for the existence of the bistable states, as well as causing the transitions between them, in contrast to the conventional picture of bistability in which the role of the noise is simply to induce transitions. A distinguishing feature of these noise-induced bistable states is the presence of a critical system size, $N_c$, above which bistability does not occur. Evidence for the effect was first found numerically in a study of autocatalytic reactions in a cell~\cite{Togashi2001}. Subsequent analytical studies proposed that the phenomenon is due to the multiplicative nature of the noise~\cite{Ohkubo2007}, and this was later confirmed by the estimation of the critical system size, $N_c$~\cite{Biancalani2012}. The theory has been applied to the study of an enzymatic cycle~\cite{Remondini2013}. A recent and more rigorous analysis can be found in~\cite{Popovic2013}.


An experimentally testable biological system that exhibits bistability may be found in the foraging behavior of an ant colony. Here we consider a classic experiment, in which a colony of ants is exposed to two identical sources of food. The foraging ants, rather than distributing equally between the two sources instead favor only one source~\cite{Pasteels1987,Detrain2006}. After a period of time they appear to turn their attention to the other option, so that the majority of ants then start to collect their food from the other source~\cite{Detrain2006, Kirman1993}. The models initially used to explain this result were typically rather detailed~\cite{Pasteels1987}. However, Kirman~\cite{Kirman1993} observed that analogous behavior also occurs in other systems involving populations, for instance queuing~\cite{Becker1991} and stock market trading~\cite{Scharfstein1990}. This suggests a common mechanism depending only on shared properties of the different systems. It is generally agreed that the autocatalytic dynamics present in all of these systems is a key ingredient required for their bistability~\cite{Deneubourg1989, Kirman1993}. 

In this Letter we propose that the underlying mechanism for the bistability observed in the experiment described above is the same as that found in the biochemical reactions previously mentioned~\cite{Togashi2001,Ohkubo2007}. To study this, we use a simple model of autocatalytic recruitment and review the estimation of the critical system size, $N_c$, using stationary analysis, for our system. However, the expression obtained for $N_c$ is not easy to experimentally test in our system. We therefore extend our analysis to study the time-dependent behavior of the system, by calculating the mean switching time between the two bistable states for different population sizes. This provides a means to measure $N_c$ experimentally and can be used to test our hypothesized mechanism for bistability.
 
Our model consists of a colony of $N$ ants collecting food from two identical sources, labeled $1$ and $2$. Ants which collect food from source $1$ are denoted by $X_1$ and those which collect food from source $2$ by $X_2$. The fraction of ants which choose source $i$ is denoted by $x_i$, $i=1,2$. An ant collecting food from one source can be recruited by an ant collecting food from the other. The recruitment of ants is thus autocatalytic, in that the more ants collecting from any particular source, the higher the rate of recruitment to that source. An ant may also spontaneously choose to use the other source. We may summarize the model through the following reaction scheme: 
\be \label{react}
	\begin{split}
	& X_1 + X_2 \xrightarrow{r} 2 X_1,\quad X_2 + X_1 \xrightarrow{r} 2 X_2, \\ 
	& X_2 \xrightarrow{\epsilon} X_1, \quad X_1 \xrightarrow{\epsilon} X_2.
	\end{split}
\ee
This model is already known in the context of chemical reactions~\cite{Ohkubo2007}, obtained as a simplification of the Togashi-Kaneko scheme~\cite{Togashi2001}. Ant recruitment is dominant so that $0 < \epsilon \ll r$, and we assume $r=1$ without loss of generality by noting that $\epsilon$ may always be rescaled, as discussed in the supplementary material (SM). We note that the number of ants is conserved so that $x_1+x_2=1$ for all time, and hence the system is fully described by a single independent variable.

To fully specify the model we now give the probability of transition, $T(a | b)$ from state $b$ to state $a$. Invoking mass action~\cite{Kampen2007}
\be \label{trates}
	\begin{split}
	& T_1 \equiv T(x_1 + \frac{1}{N},\, x_2 - \frac{1}{N} | x_1, x_2) = r x_1 x_2 + \epsilon x_2,\\
	& T_2 \equiv T(x_1 - \frac{1}{N},\, x_2 + \frac{1}{N} | x_1, x_2) = r x_1 x_2 + \epsilon x_1.
	\end{split}
\ee
We use the transition rates to write down the master equation for the probability density function (PDF), $P(x_1, x_2, t)$~\cite{Kampen2007}:
 \be \label{me}
 \begin{split}
	\partial_t P(x_1, x_2, t) =& \quad \mathclap{\sum_{(x'_1\ne x_1,x'_2 \ne x_2)}} \quad \left[ T(x_1,x_2 | x'_1,x'_2) P(x'_1,x'_2,t) \right.\\
	 &{} \left. -  T(x'_1,x'_2 | x_1,x_2) P(x_1,x_2,t) \right]. 
	 \end{split}
\ee

The scheme of reactions~\eqref{react} was simulated using the Gillespie algorithm~\cite{Gillespie1977} and a typical time series for $z=x_1-x_2$ is shown in Fig.~\ref{fig:zseries}. Regardless of the initial condition, the system settles into one of the steady states $z\approx\pm 1$, indicating that the majority of ants favor one food source. After some time, the system then switches to the other state, $z\approx\mp 1$, where the majority of ants favor the other source. 

Unlike other forms of bistability (for example, a Brownian particle in a double-well potential~\cite{Gardiner2009}), this type of bistability cannot be understood from the fixed points of the corresponding deterministic equations. Indeed, if we take the limit $N\rightarrow \infty$ \cite{Kampen2007} to eliminate stochastic effects, we obtain the equation $\dot z = - 2 \epsilon z$ (see SM). This equation has a unique stable fixed point at $z^* = 0$, which is not seen in simulations of the full system. Thus the bistability observed in the stochastic system is not reflected in the deterministic equations.
\begin{figure}[htpc!]
\includegraphics[width = 0.95\columnwidth]{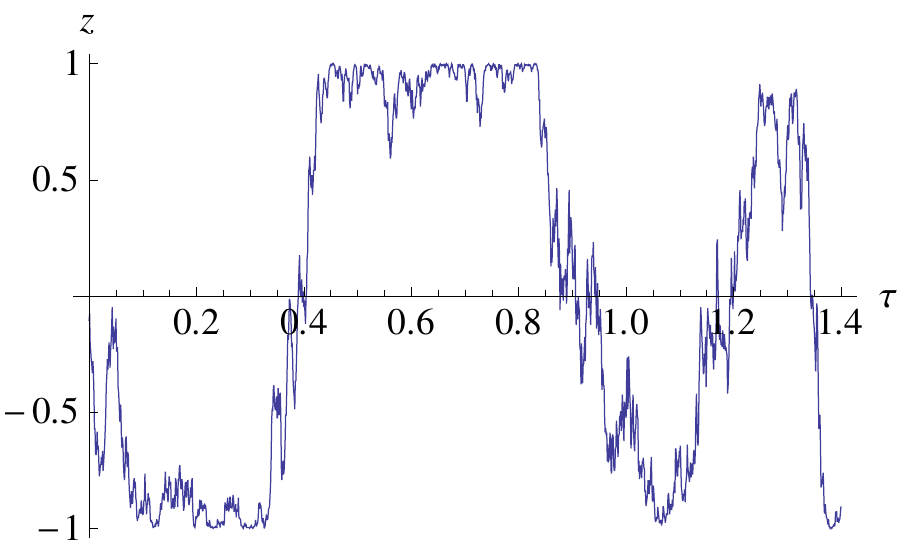} \\
\caption{(Color online) Snapshot of the time series for $z$, obtained with stochastic simulations of the scheme of reactions~\eqref{react}. Parameter values: $\epsilon=1/500$ and $N=250$. Time is expressed in units of $\tau = 2 \epsilon t/ N $. \label{fig:zseries}}
\end{figure}

To understand the origin of the bistability, we expand the master equation \eqref{me} in powers of the inverse population size, $N^{-1}$ (see SM). After rescaling time, $2 \epsilon t/ N = \tau$, we find that the system is approximated by the following stochastic differential equation (SDE) \cite{McKane2013}:
\be \label{zeq}
	z' = -z + \sqrt{\frac{N_c}{N}} \sqrt{1 + 2 \epsilon -z^2}\, \eta(\tau),
\ee
where $N_c \equiv 1/\epsilon$ and $\eta(\tau)$ is Gaussian white noise with zero mean and correlator $\langle \eta(\tau) \eta(\tau') \rangle = \delta(\tau - \tau')$. As shown in~\cite{Biancalani2012}, Eq.~\eqref{zeq} underlies a broad class of systems featuring an autocatalytic network and a slow linear reaction. The variable $z=x_1-x_2$ ranges over the interval $[-1,1]$, whose extrema correspond to all ants collecting food from a single source. Equation~\eqref{zeq} for $\epsilon=0$ is equivalent to the Wright-Fisher model with mutation, under the change of variable $x = (1+z)/2$ \cite{ewens2004mathematical}. 

We see from Eq.~\eqref{zeq} that the strength of the intrinsic system noise is proportional to $\sqrt{1+2\epsilon-z^2}$. The noise therefore has maximum strength at the deterministic steady state $z = z^*=0$, pushing the system away from this point and towards $z = \pm \sqrt{1+2\epsilon}$. Since $z$ is defined in the interval $[-1,1]$ the system cannot cross these boundaries. Bistability originates from the dependence of the noise strength on the variable $z$. At $z = \pm 1$ the noise term is at a minimum, whilst the deterministic term $-z$ attracts the system back towards $z^*$. As the trajectory leaves $z=\pm 1$ the noise term regains strength and once again kicks the system towards one of the bistable steady states $z=\pm 1$. These combined effects are seen in the dynamics of Fig.~\ref{fig:zseries}.

A distinguishing characteristic of noise-induced bistable states is the existence of a critical system size, above which bistability ceases to occur. This should be contrasted with the bistability in which the system moves between two fixed points due to the presence of noise, where varying the noise strength merely affects the characteristic time spent in each bistable state. We may therefore predict that if the bistable states are noise-induced then there should exist a critical population size above which the behavior ceases to occur. 

As shown in previous studies~\cite{ewens2004mathematical, Ohkubo2007, Biancalani2012, Remondini2013, Popovic2013}, the transition between the regime which shows bistable behavior and the one that does not, can be understood from the Fokker-Planck equation corresponding to Eq.~\eqref{zeq}. Taking $\partial_t P = 0$ and imposing zero-flux boundary conditions at $z=\pm 1$~\cite{Gardiner2009}, we obtain the stationary probability distribution
\be \label{psz}
	P_s(z) = \frac{\mathcal C_0}{\left(1+2\epsilon-z^2\right)^{1-\frac{N}{N_c}}},
\ee 
where $\mathcal C_0$ is a normalisation constant, found by requiring that the integral of $P_s(z)$ over the interval $[-1,1]$ is unity.
\begin{figure}[htpc!]
\begin{center}
\includegraphics[width = 0.95\columnwidth]{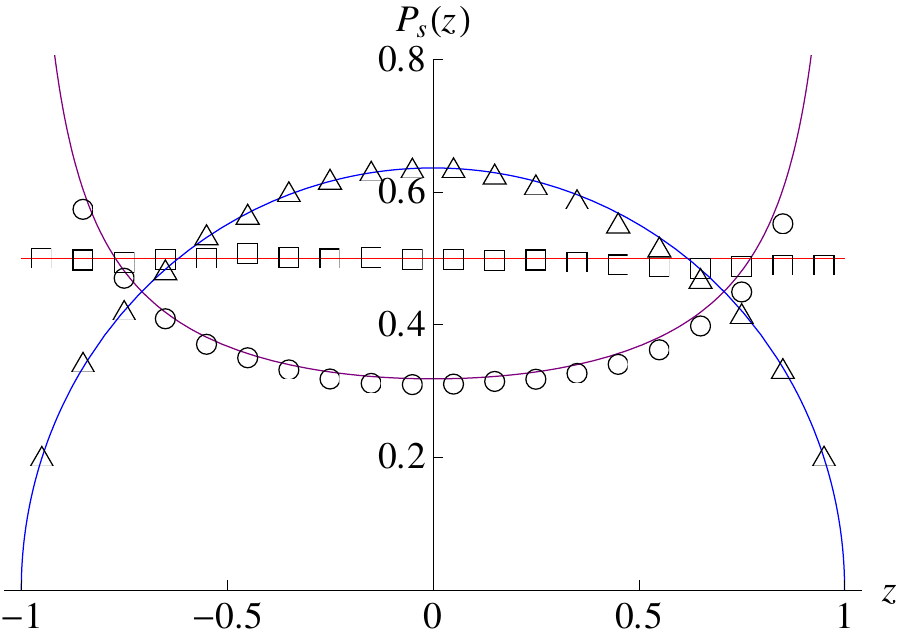} \\
\caption{(Color online) Equation \eqref{psz} (solid colors) is compared against simulations of scheme \eqref{react} (symbols). Simulations are obtained by taking the normalised histogram of a time series of length $\tau = 2.5 \times 10^9$. We have used $\epsilon = 10^{-3}$ and $N=1500$ (blue line, triangles), $N = N_c \equiv 1000$ (red line, squares) and $N=500$ (purple line, circles). \label{fig:psz}}
\end{center}
\end{figure}

The stationary distribution predicts the normalised long-time frequency histogram of $z$ and is plotted against simulation data in Fig.~\ref{fig:psz} for different population sizes. For $N < N_c$, $P_s(z)$ has a U-shape, diverging at $z=\pm \sqrt{1+2\epsilon}$. Below the critical population size, the system therefore spends most of the time close to the bistable states. In contrast, for $N > N_c$, the steady state distribution, $P_s(z)$ has an inverted U-shape, centred on the deterministic fixed point $z = z^* = 0$. This latter regime is the only one that is captured by the linear noise approximation technique (the van Kampen expansion)~\cite{McKane2013,wallace2012linear,Kampen2007}.

To estimate the critical population size requires knowledge of the parameters $r$ and $\epsilon$ (recall that we set $r=1$ by rescaling $\epsilon$). However, these reaction constants are difficult to measure experimentally. An alternative way to estimate $N_c$ is provided by calculating the time taken for the system to move from one bistable state ($z=-1$, say) to the other ($z=1$). This time is a stochastic variable whose mean (over many realizations) is denoted by $\mathcal T_\epsilon$. Using Eq.~\eqref{zeq} we may find this mean switching time~\cite{Gardiner2009} (see the SM for details). In the rescaled time variable, $\tau$, this is given by
\be \label{Teps}
	\begin{split}
	\mathcal T_\epsilon &= \frac{4N  }{(1+ 2\epsilon)N_c } \,{}_2 F_1\left(\frac{1}{2},1-\frac{N}{N_c}; \frac{3}{2}; \frac{1}{1+ 2\epsilon }\right) \times\\
	& \quad {}_2 F_1\left(\frac{1}{2},\frac{N}{N_c}; \frac{3}{2};\frac{1}{1+ 2\epsilon }\right),
	\end{split}
\ee
where the function ${}_2 F_1$ is the hypergeometric function~\cite{Abramowitz1965}. Equation~\eqref{Teps} agrees with simulations of the reaction scheme \eqref{react} only for $N$ in the neighborhood of $N_c$ (Fig.~\ref{fig:meteps}) and for $N>N_c$ (this latter result is not shown). Results are shown for different values of $\epsilon$ using different symbols. Note that for small $N$ the simulation results merge so that the mean time is independent of $\epsilon$. Since time was rescaled by $\epsilon$, however, an $\epsilon$ dependence is retained in the definition of $\tau$.
\begin{figure}[htpc!]
\begin{center}
\includegraphics[width = 0.9\columnwidth]{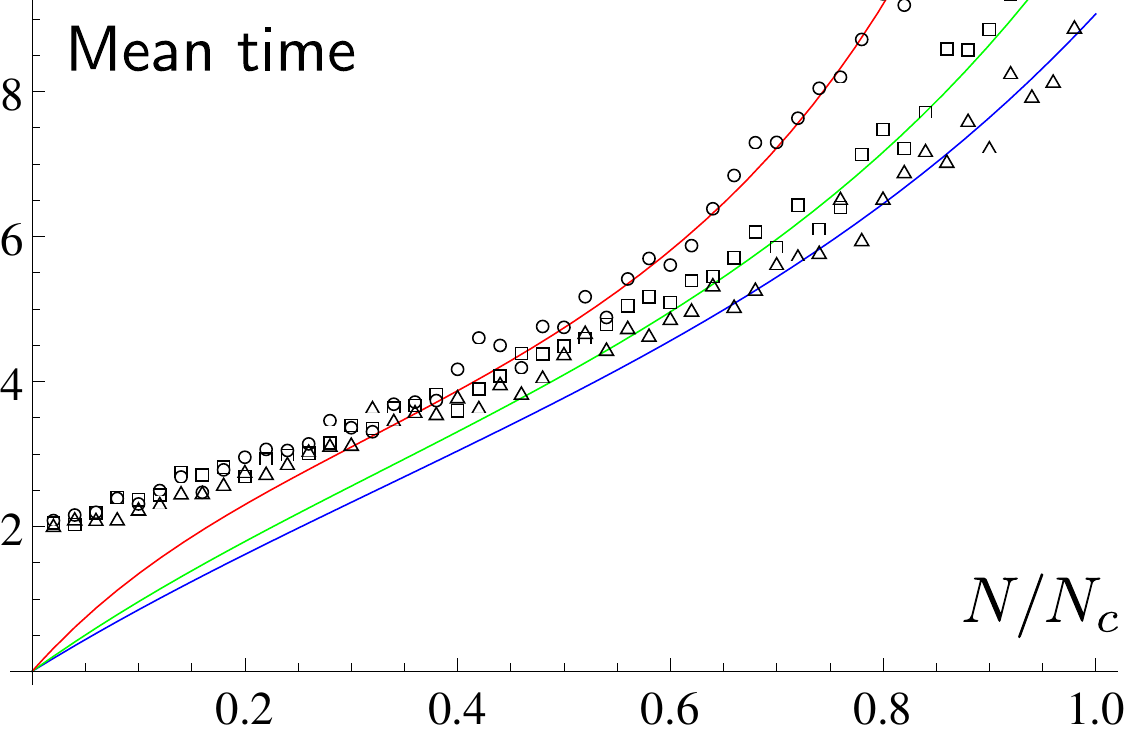} \\
\caption{(Color online) Equation~\eqref{Teps} (solid lines) is compared against stochastic simulations (symbols). Parameter used: $\epsilon = 1/50$ (blue line, triangles), $\epsilon = 1/100$ (green line, squares) and $\epsilon = 1/2000$ (red line, circles). Each symbol has been obtained by averaging over 500 simulations. \label{fig:meteps}}
\end{center}
\end{figure}

At small population sizes, as the simulation results become independent of $\epsilon$, Eq.~\eqref{Teps} breaks down and does not capture the system behavior. The failure of Eq.~\eqref{Teps} in this regime is due to assumptions made in the derivation of Eq.~\eqref{zeq}, which is no longer representative of the system at small population sizes. Instead the terms neglected in the expansion of the master equation must be retained.

Indeed, in our derivation, the noise strength in Eq.~\eqref{zeq} diverges as $N\to 0$, so that the time taken to move from one bistable state to the other shrinks to zero. In contrast, the simulated switching times do not go to zero as $N\to 0$. However, we see from Fig.~\ref{fig:metins} that the range of $N$ where our prediction holds differs for different values of $\epsilon$. The agreement improves for smaller $\epsilon$, suggesting that the limiting value of Eq.~\eqref{Teps} as $\epsilon \to 0$ may capture the system dynamics at small population sizes.
\begin{figure}[htpc!]
\begin{center}
\includegraphics[width = 0.9\columnwidth]{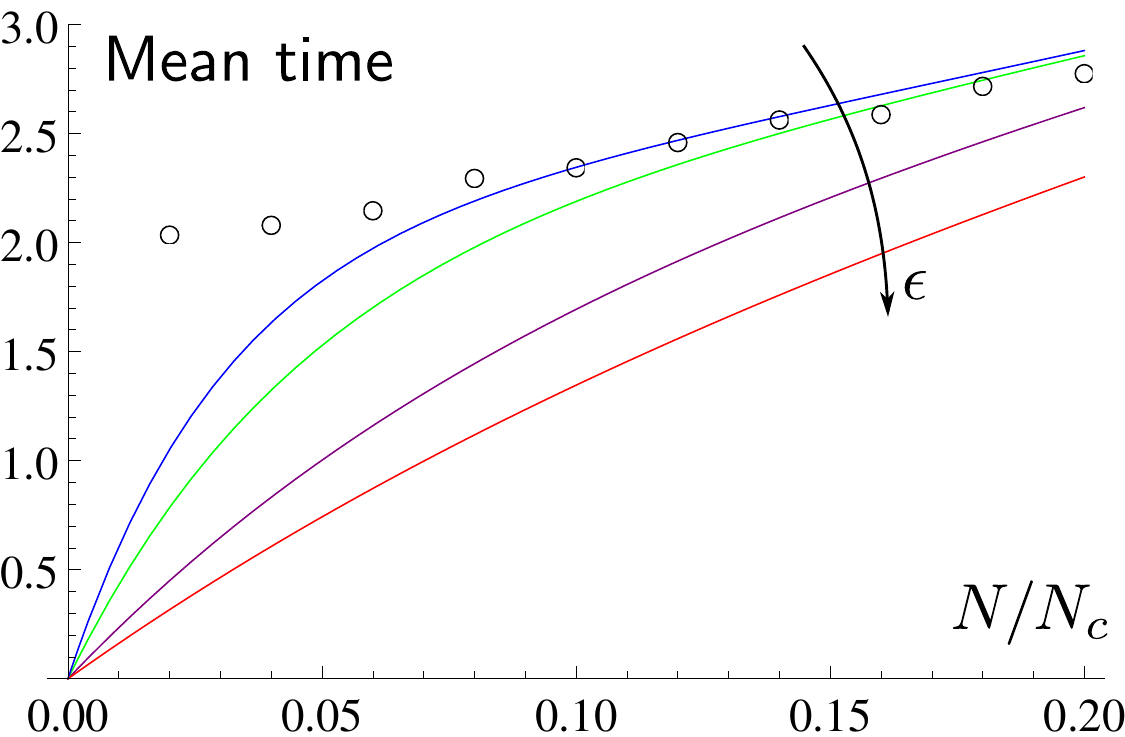} \\
\caption{(Color online) Equation~\eqref{Teps} (solid lines) is compared against stochastic simulations of the mean time for $\epsilon=1/100$ (circles). Each circle has been obtained from 2500 averages.  Parameter used for the analytical formulas: $\epsilon = 1/2000$ (red), $\epsilon = 10^{-5}$ (purple), $\epsilon = 10^{-10}$ (green) and $\epsilon=10^{-15}$ (blue). \label{fig:metins}}
\end{center}
\end{figure}

Taking $\epsilon\to 0$ (see SM), Eq.~\eqref{Teps} reduces to:
\be \label{met0}
	\mathcal T_0 =  2 \pi \frac{N}{N_c-2 N} \cot\left(\pi \frac{N}{N_c}\right).
\ee 
Equation~\eqref{met0} agrees well with simulation data for small population sizes (Fig.~\ref{fig:met0}). Since the mean switching time depends strongly on $\epsilon$ for larger population sizes (Fig.~\ref{fig:meteps}), we do not expect $\mathcal T_0$ to accurately predict the simulation data for larger $N$. Indeed, as $N\to N_c$, Eq.~\eqref{met0} diverges and thus does not capture the behavior of the system (see SM). 
\begin{figure}[htpc!]
\begin{center}
	\includegraphics[width = 0.9\columnwidth]{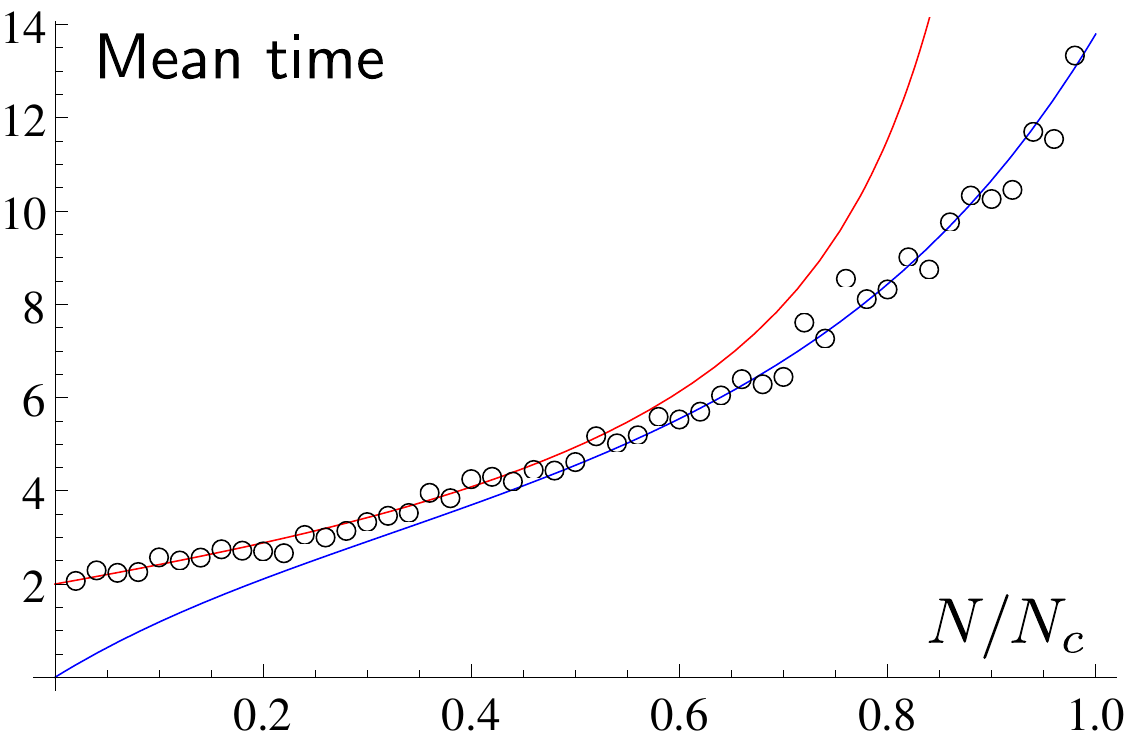} \\
\caption{(Color online) Solid lines: Eq.~\eqref{met0} (red); Eq.~\eqref{Teps} (blue) with $\epsilon=1/500$. Circles refer to stochastic simulations of the mean time (as in Fig.~\ref{fig:meteps}) with $\epsilon=1/500$.\label{fig:met0}}
\end{center}
\end{figure}

Thus we have found two expressions for the mean time to move from one bistable state to the other. Equation \eqref{Teps} is valid for larger population sizes and captures the dependence of the system on $\epsilon$ in this regime. Equation \eqref{met0} is valid for small population sizes and does not have any explicit dependence on $\epsilon$. These equations may be used to estimate both $\epsilon$ and the critical population size, $N_c$. To facilitate this estimation we first linearize Eq.~\eqref{met0} for small $N$ to obtain $\mathcal T_0 \approx 4 N/N_c + 2$.

Since $\mathcal T_0$ is measured in units of $\tau = 2\epsilon t/ N$, and $\epsilon$ is unknown, we may plot experimental results for $t/N$ and observe that we would expect to obtain a straight line for small values of $N$. The $y$-intercept is then given by $\epsilon^{-1}$, whilst the gradient will be $2/ ( N_c \epsilon )$. The value obtained for $\epsilon$ may then be checked by taking larger population sizes and using Eq.~\eqref{Teps}. Note, however, that the value of $\epsilon$ found is the ratio of the two reaction constants, $r$ and $\epsilon$, since $\epsilon$ has been rescaled in order to take $r=1$.

In this Letter we have presented a way to experimentally determine the critical population size in a system with noise-induced bistable states. Using time-dependent analysis, we have investigated the mean time taken for the system to move between the two bistable states and found that two regimes exist. For small population sizes, the mean switching time is independent of $\epsilon$ and Eq.~\eqref{met0} is representative of the system behavior. Conversely, for large population sizes the value of $\epsilon$ becomes important and we must use Eq.~\eqref{Teps}. The mean switching time is an experimentally measurable quantity that may be used to confirm or reject the hypothesis that noise-induced bistable states may explain the empirical results seen in the experiments on ant foraging..

The analysis may be further extended by considering the full distribution of times to move between the bistable states, rather than using only the mean time. In this way it would be possible to assess any skewness of the distribution and determine how representative the mean time is of the full distribution.

Our results do not only apply to the model described here, as Eq.~\eqref{zeq} is the reduced one-dimensional equation for many stochastic systems, such as the Togashi-Kaneko model~\cite{Biancalani2012}. We believe that the mechanism for noise-induced bistability, in which the changing noise strength at different system states leads to substantially different behavior from the deterministic approximation, will be applicable to a wide variety of systems.

\vspace{10pt}
\begin{acknowledgments}
TB acknowledges partial financial support from the EPSRC (UK) and LD was supported under EPSRC grant EP/H02171X.
\end{acknowledgments}

\onecolumngrid

\appendix

\subsection{The derivation of the equation for the $z$ variable}
\noindent The model is defined by the two transition rates:
\be \label{S:trates}
	\begin{split}
	& T_1 \equiv T(x_1 + \frac{1}{N},\, x_2 - \frac{1}{N} | x_1, x_2) = r x_1 x_2 + \epsilon x_2,\\
	& T_2 \equiv T(x_1 - \frac{1}{N},\, x_2 + \frac{1}{N} | x_1, x_2) = r x_1 x_2 + \epsilon x_1.
	\end{split}
\ee
We rewrite the master equation,
\be \label{S:me}
	\partial_t P(x_1, x_2, t) = \sum_{(x'_1\ne x_1,x'_2 \ne x_2)} \left[ T(x_1,x_2 | x'_1,x'_2) P(x'_1,x'_2,t) - T(x'_1,x'_2 | x_1,x_2)  P(x_1,x_2,t) \right],
\ee
using the step operators, $\varepsilon_i^{\pm}$, which represent the creation or destruction of a molecule of species $X_i$ ($i=1,2$). Taylor expanding in $1/N$, the inverse of the population size:
\be \label{S:expansion}
	\varepsilon_i^{\pm} f(x_i) = f(x_i \pm \frac{1}{N}) \approx \left(1 \pm \frac{1}{N}\partial_{x_i} + \frac{1}{2 N^2}\partial_{x_i}^2\right) f(x_i),
\ee
where $f(x_i)$ is a general function of the fraction of the $i$-th species, $x_i$. The master equation~\eqref{S:me} can be approximated using Eq.~\eqref{S:expansion} to give
\be
	\begin{split}
		\partial_t P(x_1,x_2,t) &= \left[\left(\varepsilon_1^-\varepsilon_2^+ -1 \right) T_1 + \left(\varepsilon_1^+\varepsilon_2^- -1 \right)T_2 \right] P(x_1,x_2,t) \\
		&\approx \left[ \frac{1}{N} \left(\partial_{x_2} - \partial_{x_1}\right) T_1 + \frac{1}{N} \left(\partial_{x_1} - \partial_{x_2}\right) T_2 + \frac{1}{2 N^2} \left(\partial_{x_1} - \partial_{x_2}\right)^2 \left(T_1+T_2\right) \right] P(x_1,x_2,t),
	\end{split}
\ee
neglecting terms of $\mathcal{O}(1/N^3)$.

Rescaling time by $t/N \to t$ and inserting the expressions of the transition rates~\eqref{S:trates} gives the Fokker-Planck equation
\be
	\partial_t  P(x_1,x_2,t) =  \left[ - \partial_{x_1} \mathcal A_1 - \partial_{x_2} \mathcal A_2 + \frac{1}{2 N} \sum_{i,j=1}^2 \partial_{x_i} \partial_{x_j} \mathcal B_{ij} \right] P(x_1,x_2,t),
\ee
where $\mathcal A_1 = - \mathcal A_2 = \epsilon (x_2-x_1)$ and $\mathcal B_{ij} = (2 r x_1 x_2 + \epsilon (x_1+x_2))(-1)^{i+j}$. This is equivalent to the following system of SDEs~\cite{Gardiner2009} in which the noises have zero mean:
\be \label{S:xieq}
	\begin{split}
	 \dot x_1 &= \mathcal A_1 + \frac{1}{\sqrt N}\xi_1, \quad \dot x_2 = \mathcal A_2 + \frac{1}{\sqrt N}\xi_2, \qquad \langle \xi_i(t) \xi_j(t') \rangle =\mathcal B_{ij} \delta(t-t').
	\end{split}
\ee

We make the transformation $\xi_i = \sum_{j=1}^2 \mathcal G_{ij} \eta_j$, where $\mathcal G_{ij} = (-1)^{i+j+1}\sqrt{2rx_1 x_2 + \epsilon(x_1+x_2)}/\sqrt 2 $. Hence the new noises are delta-correlated, that is, $\langle \eta_i(t) \eta_j(t') \rangle = \delta_{ij}\,\delta(t-t')$. This can be proved using the expression of the correlator for $\xi_i$ and the fact that $\mathcal B = \mathcal G \mathcal G^T$. System~\eqref{S:xieq} then becomes:
\be \label{S:xieq2}
	 \dot x_1 = \epsilon (x_2-x_1) + \sqrt{\frac{2 r x_1 x_2 + \epsilon (x_1 + x_2)}{2 N}} \left(\eta_2 -\eta_1\right) = - \dot x_2.
\ee

We now introduce new variables, $w=x_1+x_2$ and $z=x_1-x_2$, which satisfy equations obtained by summing and subtracting the equations for $x_1$ and $x_2$:
\be \label{S:wzeq}
	 \dot w = 0, \quad \dot z = -2 \epsilon z + \sqrt{\frac{1}{N}} \sqrt{r \left(w^2-z^2\right)+2 \epsilon w} \left( \eta_2 - \eta_1 \right).
\ee
The $z$ equation can be simplified as follows. First, we use the sum rule for Gaussian variables, so that $\eta_1 - \eta_2 = \sqrt 2 \eta$, where $\eta$ is normalised Gaussian white noise~\cite{Gardiner2009}. Then, we rescale time by $2 \epsilon t \to t$. Note that the coefficient which multiplies the noise scales with a square root law, as expected~\cite{Gardiner2009}. The overall time scaling is given by $\tau = 2\epsilon t/{N}$ and we obtain
\be \label{S:wzeqtau}
	w' = 0, \quad z' = - z + \sqrt{\frac{r}{N\epsilon}} \sqrt{\left(w^2-z^2\right)+ 2 \frac{\epsilon}{r} w}\, \eta(\tau),
\ee
where the prime sign indicates the time derivative with respect to $\tau$. Without loss of generality we set $r=1$, since we may rescale $\epsilon$ to absorb $r$. Since the transition rates~\eqref{S:trates} do not alter the total number of ants, $N$ and $w$ are conserved quantities with $w=1$. Hence
\be \label{S:zeq}
	z' = -z + \sqrt{\frac{N_c}{N}} \sqrt{1-z^2+ 2 \epsilon }\, \eta(\tau).
\ee
where $N_c = 1/\epsilon$.

\subsection{The mean switching time} 
We wish to find the mean time for the system to leave $z=-1$ and reach $z=1$. To derive this quantity we consider the mean time, $\langle \mathsf T_\epsilon \rangle$, for a system starting at $z$ to leave the interval $[-1,1]$. This is derived from $G(z,t)$, the density of probability that a system beginning at $z$ has not left the interval $[-1,1]$ by time $t$. Then $G(z,t)$ satisfies the backward Fokker-Planck equation corresponding to Eq.~\eqref{S:zeq}~\cite{Gardiner2009}
\be \label{S:Geq}
	\D{G}{t} = -z \D{G}{z} + \frac{1}{2\lambda} \left(1-z^2 + 2\epsilon\right) \DD{G}{z},
\ee 
where $\lambda = N/N_c$, with a reflecting boundary condition at $z=-1$ and an absorbing boundary condition at $z=1$. Now the probability density function for the system beginning at $z$ and reaching the boundary at $z=1$ at time $t$ (where it is thus removed from the interval) is given by $-\partial_t G$~\cite{Gardiner2009}. Hence the mean switching time is given by
\be
	\langle \mathsf T_\epsilon \rangle = -\int_0^\infty t \partial_t G(z,t) \dd t = \int_0^\infty G(z,t) \dd t ,
\ee
assuming $G(z,t)$ is well behaved as $t \rightarrow \infty$. Integrating Eq.~\eqref{S:Geq} over $t$, we obtain
\be \label{S:meteq}
	\frac{1}{2\lambda} \left(1-z^2 + 2\epsilon\right) \langle \mathsf T_\epsilon \rangle''-z \langle \mathsf T_\epsilon \rangle'+1 = 0,
\ee 
since the system must start in the interval $[-1,1]$ so that
\be
	\int_0^\infty \partial_t G(z,t) \dd t = -G(z,0) = -1.
\ee
We may solve Eq.~\eqref{S:meteq} by first writing it as
\be\label{S:IF}
	\frac{d }{dz}\,\left((1+2\epsilon - z^2)^\lambda \langle \mathsf T_\epsilon \rangle'\right) = -2\lambda (1+2\epsilon - z^2)^{\lambda-1}.
\ee
To integrate the right hand side we need the following integral
\be \label{S:I}
	\begin{split}
	I_1(\mu) &= \int (1+2\epsilon - z^2)^\mu dz,\\
	&= \frac{z(1-z^2+2\epsilon)^{1+\mu}}{1 + 2 \epsilon} \,{}_2 F_1\left(1, \frac{3}{2} + \mu, \frac{3}{2}, \frac{z^2}{1+2\epsilon}\right),
	\end{split}
\ee	
where ${}_2 F_1$ is the hypergeometric function~\cite{Abramowitz1965}. Equality~\eqref{S:I} can be seen by expanding $(1+2\epsilon - z^2)^\mu$ as the binomial series, integrating term-by-term and using the series definition for the hypergeometric function. Integrating both sides of Eq.~\eqref{S:IF}:
\be
	\langle \mathsf T_\epsilon \rangle' = -\frac{2\lambda z}{1 + 2 \epsilon} \,{}_2 F_1\left(1, \frac{1}{2} + \lambda, \frac{3}{2}, \frac{z^2}{1+2\epsilon}\right)+ C_1\, (1+2\epsilon - z^2)^{-\lambda},
\ee
where $C_1$ is an integration constant. Integrating again, we obtain
\be
	\langle \mathsf T_\epsilon \rangle = I_2 + C_1 I_1(-\lambda) + \mathcal C_2,
\ee
where
\be
	\begin{split}
	I_2 &= -\frac{2\lambda }{1 + 2 \epsilon} \int z\,{}_2 F_1\left(1, \frac{1}{2} + \lambda, \frac{3}{2}, \frac{z^2}{1+2\epsilon}\right) dz,\\
	&= - \frac{\lambda z^2}{(1+2\epsilon)} \,{}_3 F_2\left(1,1,\frac{1}{2}+\lambda;\frac{3}{2},2;\frac{z^2}{1+2\epsilon} \right),
	\end{split}
\ee
since~\cite{Slater1966}
\be 
	\int z^{\alpha-1} \,{}_2 F_1\left(a,b,c;z\right) dz = \frac{z^\alpha}{\alpha} \,{}_3 F_2\left(a,b,\alpha; c, 1+\alpha, z\right),
\ee
where ${}_p F_q(x_1,\ldots,x_q;y_1,\ldots,y_p;z)$ indicates the generalised hypergeometric function~\cite{Abramowitz1965, Slater1966}.

The final expression is therefore
\be \label{S:met}
	\langle \mathsf T_\epsilon \rangle = \mathcal C_2+z\, \mathcal C_1 \,{}_2 F_1\left(\frac{1}{2},\lambda; \frac{3}{2}; \frac{z^2}{1+2\epsilon }\right) - \frac{z^2 \lambda}{(1+2\epsilon)} \,{}_3 F_2\left(1,1,\frac{1}{2}+\lambda;\frac{3}{2},2;\frac{z^2}{1+2\epsilon} \right),
\ee
where $\mathcal C_1 = C_1 / (1+2\epsilon)$ and we have used the Euler transformation, $(1 - w)^{a + b - c} \,{}_2 F_1 (a, b; c; w) = \,{}_2 F_1 (c - a, c - b; c; w)$, to simplify the second term.

We take a reflecting boundary condition at $z=-1$ and an absorbing boundary condition at $z=1$. Hence with the initial condition $z=-1$, we model a trajectory that begins at $z=-1$ and is stopped at $z=1$. These two boundary conditions determine the constants $\mathcal C_1$ and $\mathcal C_2$ and are given by~\cite{Gardiner2009}:
\be \label{S:meticbc}
	\langle \mathsf T_\epsilon \rangle(z=1) = 0, \quad \langle \mathsf T_\epsilon \rangle'(z=-1) = 0.
\ee  
To determine the constant $\mathcal C_2$ we use the absorbing boundary condition in Eq.~\eqref{S:met}, so that
\be \label{S:c2c1}
	\mathcal C_2 = - \mathcal C_1 \,{}_2 F_1\left(\frac{1}{2},\lambda; \frac{3}{2}; \frac{1}{1+2\epsilon}\right) + \frac{\lambda}{(1+2\epsilon)} \, {}_3 F_2\left(1,1,\frac{1}{2}+\lambda; \frac{3}{2},2;\frac{1}{1+2\epsilon} \right).
\ee
Thus, $\mathcal C_2$ is fully determined once $\mathcal C_1$ has been found.

To satisfy the reflecting boundary condition, we differentiate Eq.~\eqref{S:met}:
\be \label{S:metp}
	\langle \mathsf T_\epsilon \rangle'= \left(1-\frac{z^2}{1+2\epsilon }\right)^{-\lambda} \left[\mathcal{C}_1 - \frac{2\lambda z}{(1+2\epsilon )}  \,{}_2 F_1\left(\frac{1}{2},1-\lambda; \frac{3}{2}; \frac{z^2}{1+2\epsilon }\right)\right].
\ee
The reflecting boundary condition is satisfied if the term in the square brackets converges to zero as $z \to -1$. This yields:
\be \label{S:c1}
	\mathcal C_1 = -\frac{2\lambda }{1+2\epsilon } {}_2 F_1\left(\frac{1}{2},1-\lambda; \frac{3}{2}; \frac{1}{1+ 2\epsilon }\right).
\ee 
Inserting $\mathcal C_1$ into Eq.~\eqref{S:c2c1} leads to an expression for $\mathcal C_2$:
\be	\label{S:c2}
	\mathcal C_2 = \frac{\lambda}{(1+2\epsilon )} \left[2 \,{}_2 F_1\left(\frac{1}{2},1-\lambda;\frac{3}{2};\frac{1}{1+ 2\epsilon }\right) \,{}_2 F_1\left(\frac{1}{2},\lambda; \frac{3}{2}; \frac{1}{1+ 2\epsilon }\right)+\,{}_3 F_2\left(1,1,\frac{1}{2}+\lambda; \frac{3}{2},2;\frac{1}{1+2\epsilon}\right)\right].
\ee

We now use the expressions in Eq.~\eqref{S:c1} and Eq.~\eqref{S:c2} in Eq.~\eqref{S:met}. We also set the initial condition, $z=-1$. Hence the final formula for the mean time for the system to leave $z=-1$ and reach $z=1$ is 
\be \label{S:meteps}
	\mathcal T_\epsilon \equiv \langle \mathsf T_\epsilon \rangle(-1) = \frac{4 \lambda  }{1+ 2\epsilon } \,{}_2 F_1\left(\frac{1}{2},1-\lambda; \frac{3}{2}; \frac{1}{1+ 2\epsilon }\right) \,{}_2 F_1\left(\frac{1}{2},\lambda; \frac{3}{2};\frac{1}{1+ 2\epsilon }\right).
\ee
Note that as $\epsilon\to 0$ the expression of the mean time may be written as
\BE
	\mathcal T_0 \equiv \langle \mathsf T_0 \rangle(-1) = 4\lambda \frac{ \Gamma \left(\frac{3}{2}\right)^2 \Gamma\left(\frac{\lambda}{2}\right) \Gamma \left(1-\frac{\lambda}{2}\right)}{\Gamma(1)^2 \Gamma\left(\frac{\lambda + 1}{2}\right) \Gamma\left(\frac{3-\lambda}{2}\right) } = \pi \lambda \frac{ \Gamma\left(\frac{\lambda}{2}\right) \Gamma \left(1-\frac{\lambda}{2}\right)}{\Gamma\left(\frac{\lambda + 1}{2}\right) \Gamma\left(\frac{3-\lambda}{2}\right) },
\EE
since~\cite{Abramowitz1965}
\BE
	{}_2 F_1(a,b,c,1) = \frac{\Gamma(c)\Gamma(c-a-b)}{\Gamma(c-a)\Gamma(c-b)}, \quad \Gamma\left(\frac{3}{2}\right) = \frac{\sqrt{\pi}}{2}, \quad \Gamma(1)=1.
\EE
But now~\cite{Abramowitz1965} $\Gamma(z)\Gamma(1-z) = \pi \, \csc(\pi z)$, $\Gamma(z+1) = z \Gamma(z)$ and $\csc(z+\pi/2) = \sec(z)$ so that
\be \label{S:met0}
	\mathcal T_0  = \pi \frac{2\lambda }{1-2\lambda } \cot\left(\pi  \lambda\right).
\ee

Equation~\eqref{S:met0} diverges as $\lambda\to 1$. To understand why, we classify the boundaries of our original SDE (Eq.~\eqref{S:zeq}). In fact, SDEs with multiplicative noise may exhibit pathological behaviour that can be detected (or ruled out) by calculating three integrals, $\mathcal L_i$, for $i=1,2,3$~\cite{Kampen2007}. For Eq.\eqref{S:zeq}, the first of these integrals reads:
\be \label{S:L1}
	\mathcal L_1(z) = \int_0^{z} dy\,\left(1-\frac{y^2}{1+2\epsilon}\right)^{\lambda} = z\, {}_2F_1\left(\frac{1}{2},\lambda ,\frac{3}{2},\frac{z^2}{1+2\epsilon}\right).
\ee
This integral determines whether or not the point $z$ can be reached by the stochastic trajectory. If the $\mathcal L_1(z)=\infty$, then the trajectory cannot reach $z$, and $z$ is called a natural repulsive boundary. Equation \eqref{S:L1} diverges as $\lambda\to 1$ if $z^2 / (1+2\epsilon) = 1$. Thus for $\epsilon=0$ the point $z=1$ is not reachable for $\lambda=1$ and the mean time to reach the boundary diverges. In contrast, for $\epsilon>0$ the $z = 1/(1+2\epsilon)$ is not contained in $[-1,1]$, the interval of definition for $z$.

This concludes the analytical treatment of the mean time. As a final remark, note that if one wants to derive Eq.~\eqref{S:met0} starting from Eq.~\eqref{S:meteq} for $\epsilon=0$, more care is required in formulating the boundary conditions. In fact, Eq.~\eqref{S:meteq} for $\epsilon=0$ becomes singular as $z \to -1$, in the sense that the coefficient which multiplies the second derivative vanishes in that limit. The boundary conditions must therefore be modified to~\cite{Gardiner2009}
\be 
	\langle \mathsf T_0 \rangle(z=1) = 0, \quad \lim_{z \to -1} \left(1-z^2 \right) \langle \mathsf T_0 \rangle'(z) = 0,
\ee
representing the absorbing boundary at $z=1$ and reflecting boundary at $z=-1$. The calculation may then be carried out analogously to the previous derivation. However, we find that the reflecting boundary condition is inherently satisfied regardless of the choice of $\mathcal C_1$. To determine this constant, we must require an additional condition, namely that the derivative of the mean time remain finite as $z\to -1$. This condition is automatically satisfied when $\epsilon$ is non-zero (as can be seen from Eq.~\eqref{S:metp}). Using all these conditions one can derive Eq.~\eqref{S:met0}.

\bibliographystyle{apsrev4-1}
\bibliography{ants}

\begin{thebibliography}{21}%
\makeatletter
\providecommand \@ifxundefined [1]{%
 \@ifx{#1\undefined}
}%
\providecommand \@ifnum [1]{%
 \ifnum #1\expandafter \@firstoftwo
 \else \expandafter \@secondoftwo
 \fi
}%
\providecommand \@ifx [1]{%
 \ifx #1\expandafter \@firstoftwo
 \else \expandafter \@secondoftwo
 \fi
}%
\providecommand \natexlab [1]{#1}%
\providecommand \enquote  [1]{``#1''}%
\providecommand \bibnamefont  [1]{#1}%
\providecommand \bibfnamefont [1]{#1}%
\providecommand \citenamefont [1]{#1}%
\providecommand \href@noop [0]{\@secondoftwo}%
\providecommand \href [0]{\begingroup \@sanitize@url \@href}%
\providecommand \@href[1]{\@@startlink{#1}\@@href}%
\providecommand \@@href[1]{\endgroup#1\@@endlink}%
\providecommand \@sanitize@url [0]{\catcode `\\12\catcode `\$12\catcode
  `\&12\catcode `\#12\catcode `\^12\catcode `\_12\catcode `\%12\relax}%
\providecommand \@@startlink[1]{}%
\providecommand \@@endlink[0]{}%
\providecommand \url  [0]{\begingroup\@sanitize@url \@url }%
\providecommand \@url [1]{\endgroup\@href {#1}{\urlprefix }}%
\providecommand \urlprefix  [0]{URL }%
\providecommand \Eprint [0]{\href }%
\@ifxundefined \urlstyle {%
  \providecommand \doi  [0]{\begingroup \@sanitize@url \@doi}%
  \providecommand \@doi [1]{\endgroup \@@startlink {\doibase
  #1}doi:\discretionary {}{}{}#1\@@endlink }%
}{%
  \providecommand \doi  [0]{doi:\discretionary{}{}{}\begingroup
  \urlstyle{rm}\Url }%
}%
\providecommand \doibase [0]{http://dx.doi.org/}%
\providecommand \Doi [0]{\begingroup \@sanitize@url \@Doi }%
\providecommand \@Doi  [1]{\endgroup\@@startlink{\doibase#1}\@@Doi}%
\providecommand \@@Doi [1]{#1\@@endlink}%
\providecommand \selectlanguage [0]{\@gobble}%
\providecommand \bibinfo  [0]{\@secondoftwo}%
\providecommand \bibfield  [0]{\@secondoftwo}%
\providecommand \translation [1]{[#1]}%
\providecommand \BibitemOpen [0]{}%
\providecommand \bibitemStop [0]{}%
\providecommand \bibitemNoStop [0]{.\EOS\space}%
\providecommand \EOS [0]{\spacefactor3000\relax}%
\providecommand \BibitemShut  [1]{\csname bibitem#1\endcsname}%
\bibitem [{\citenamefont {Gardiner}(2009)}]{Gardiner2009}%
  \BibitemOpen
  \bibfield  {author} {\bibinfo {author} {\bibfnamefont {C.~W.}\ \bibnamefont
  {Gardiner}},\ }\href@noop {} {\emph {\bibinfo {title} {{Handbook of
  Stochastic Methods for Physics, Chemistry and the Natural Sciences}}}},\
  \bibinfo {edition} {4th}\ ed.\ (\bibinfo  {publisher} {Springer},\ \bibinfo
  {address} {New York},\ \bibinfo {year} {2009})\BibitemShut {NoStop}%
\bibitem [{\citenamefont {McKane}\ and\ \citenamefont
  {Newman}(2004)}]{McKane2004}%
  \BibitemOpen
  \bibfield  {author} {\bibinfo {author} {\bibfnamefont {A.~J.}\ \bibnamefont
  {McKane}}\ and\ \bibinfo {author} {\bibfnamefont {T.~J.}\ \bibnamefont
  {Newman}},\ }\href@noop {} {\bibfield  {journal} {\bibinfo  {journal} {Phys.
  Rev. E},\ }\textbf {\bibinfo {volume} {70}},\ \bibinfo {pages} {041902}
  (\bibinfo {year} {2004})}\BibitemShut {NoStop}%
\bibitem [{\citenamefont {Gillespie}\ \emph {et~al.}(2013)\citenamefont
  {Gillespie}, \citenamefont {Hellander},\ and\ \citenamefont
  {Petzold}}]{gillespie2013perspective}%
  \BibitemOpen
  \bibfield  {author} {\bibinfo {author} {\bibfnamefont {D.~T.}\ \bibnamefont
  {Gillespie}}, \bibinfo {author} {\bibfnamefont {A.}~\bibnamefont
  {Hellander}}, \ and\ \bibinfo {author} {\bibfnamefont {L.~R.}\ \bibnamefont
  {Petzold}},\ }\href@noop {} {\bibfield  {journal} {\bibinfo  {journal} {J.
  Chem. Phys.},\ }\textbf {\bibinfo {volume} {138}},\ \bibinfo {pages} {170901}
  (\bibinfo {year} {2013})}\BibitemShut {NoStop}%
\bibitem [{\citenamefont {Togashi}\ and\ \citenamefont
  {Kaneko}(2001)}]{Togashi2001}%
  \BibitemOpen
  \bibfield  {author} {\bibinfo {author} {\bibfnamefont {Y.}~\bibnamefont
  {Togashi}}\ and\ \bibinfo {author} {\bibfnamefont {K.}~\bibnamefont
  {Kaneko}},\ }\Doi {10.1103/PhysRevLett.86.2459} {\bibfield  {journal}
  {\bibinfo  {journal} {Phys. Rev. Lett.},\ }\textbf {\bibinfo {volume} {86}},\
  \bibinfo {pages} {2459} (\bibinfo {year} {2001})}\BibitemShut {NoStop}%
\bibitem [{\citenamefont {Ohkubo}\ \emph {et~al.}(2007)\citenamefont {Ohkubo},
  \citenamefont {Shnerb},\ and\ \citenamefont {Kessler}}]{Ohkubo2007}%
  \BibitemOpen
  \bibfield  {author} {\bibinfo {author} {\bibfnamefont {J.}~\bibnamefont
  {Ohkubo}}, \bibinfo {author} {\bibfnamefont {N.}~\bibnamefont {Shnerb}}, \
  and\ \bibinfo {author} {\bibfnamefont {D.}~\bibnamefont {Kessler}},\ }\Doi
  {10.1143/JPSJ.77.044002} {\bibfield  {journal} {\bibinfo  {journal} {J. Phys.
  Soc. Jpn.},\ }\textbf {\bibinfo {volume} {77}},\ \bibinfo {pages} {044002}
  (\bibinfo {year} {2007})}\BibitemShut {NoStop}%
\bibitem [{\citenamefont {Biancalani}\ \emph {et~al.}(2012)\citenamefont
  {Biancalani}, \citenamefont {Rogers},\ and\ \citenamefont
  {McKane}}]{Biancalani2012}%
  \BibitemOpen
  \bibfield  {author} {\bibinfo {author} {\bibfnamefont {T.}~\bibnamefont
  {Biancalani}}, \bibinfo {author} {\bibfnamefont {T.}~\bibnamefont {Rogers}},
  \ and\ \bibinfo {author} {\bibfnamefont {A.~J.}\ \bibnamefont {McKane}},\
  }\href {http://pre.aps.org/abstract/PRE/v86/i1/e010106} {\bibfield  {journal}
  {\bibinfo  {journal} {Phys. Rev. E},\ }\textbf {\bibinfo {volume} {86}},\
  \bibinfo {pages} {010106(R)} (\bibinfo {year} {2012})}\BibitemShut {NoStop}%
\bibitem [{\citenamefont {Remondini}\ \emph {et~al.}(2013)\citenamefont
  {Remondini}, \citenamefont {Giampieri}, \citenamefont {Bazzani},
  \citenamefont {Castellani},\ and\ \citenamefont {Maritan}}]{Remondini2013}%
  \BibitemOpen
  \bibfield  {author} {\bibinfo {author} {\bibfnamefont {D.}~\bibnamefont
  {Remondini}}, \bibinfo {author} {\bibfnamefont {E.}~\bibnamefont
  {Giampieri}}, \bibinfo {author} {\bibfnamefont {A.}~\bibnamefont {Bazzani}},
  \bibinfo {author} {\bibfnamefont {G.}~\bibnamefont {Castellani}}, \ and\
  \bibinfo {author} {\bibfnamefont {A.}~\bibnamefont {Maritan}},\ }\Doi
  {10.1016/j.physa.2012.09.005} {\bibfield  {journal} {\bibinfo  {journal}
  {Physica A},\ }\textbf {\bibinfo {volume} {392}},\ \bibinfo {pages} {336}
  (\bibinfo {year} {2013})}\BibitemShut {NoStop}%
\bibitem [{\citenamefont {Popovic}\ and\ \citenamefont
  {McSweeney}(2013)}]{Popovic2013}%
  \BibitemOpen
  \bibfield  {author} {\bibinfo {author} {\bibfnamefont {L.}~\bibnamefont
  {Popovic}}\ and\ \bibinfo {author} {\bibfnamefont {J.}~\bibnamefont
  {McSweeney}},\ }\href {http://arxiv.org/abs/1302.1446
  http://www.arxiv.org/pdf/1302.1446.pdf} {\bibfield  {journal} {\bibinfo
  {journal} {arXiv preprint arXiv:1302.1446}} (\bibinfo {year}
  {2013})}\BibitemShut {NoStop}%
\bibitem [{\citenamefont {Pasteels}\ \emph {et~al.}(1987)\citenamefont
  {Pasteels}, \citenamefont {Deneubourg},\ and\ \citenamefont
  {Goss}}]{Pasteels1987}%
  \BibitemOpen
  \bibfield  {author} {\bibinfo {author} {\bibfnamefont {J.}~\bibnamefont
  {Pasteels}}, \bibinfo {author} {\bibfnamefont {J.}~\bibnamefont
  {Deneubourg}}, \ and\ \bibinfo {author} {\bibfnamefont {S.}~\bibnamefont
  {Goss}},\ }\href@noop {} {\bibfield  {journal} {\bibinfo  {journal}
  {Experientia Supplementum},\ }\textbf {\bibinfo {volume} {54}},\ \bibinfo
  {pages} {155} (\bibinfo {year} {1987})}\BibitemShut {NoStop}%
\bibitem [{\citenamefont {Detrain}\ and\ \citenamefont
  {Deneubourg}(2006)}]{Detrain2006}%
  \BibitemOpen
  \bibfield  {author} {\bibinfo {author} {\bibfnamefont {C.}~\bibnamefont
  {Detrain}}\ and\ \bibinfo {author} {\bibfnamefont {J.}~\bibnamefont
  {Deneubourg}},\ }\Doi {10.1016/j.plrev.2006.07.001} {\bibfield  {journal}
  {\bibinfo  {journal} {Phys. Life Rev.},\ }\textbf {\bibinfo {volume} {3}},\
  \bibinfo {pages} {162} (\bibinfo {year} {2006})}\BibitemShut {NoStop}%
\bibitem [{\citenamefont {Kirman}(1993)}]{Kirman1993}%
  \BibitemOpen
  \bibfield  {author} {\bibinfo {author} {\bibfnamefont {A.}~\bibnamefont
  {Kirman}},\ }\href {http://cadmus.eui.eu/handle/1814/16771
  http://qje.oxfordjournals.org/content/108/1/137.short} {\bibfield  {journal}
  {\bibinfo  {journal} {Q. J. Econ.},\ }\textbf {\bibinfo {volume} {108}},\
  \bibinfo {pages} {137} (\bibinfo {year} {1993})}\BibitemShut {NoStop}%
\bibitem [{\citenamefont {Becker}(1991)}]{Becker1991}%
  \BibitemOpen
  \bibfield  {author} {\bibinfo {author} {\bibfnamefont {G.~S.}\ \bibnamefont
  {Becker}},\ }\href@noop {} {\bibfield  {journal} {\bibinfo  {journal} {J.
  Polit. Econ.},\ }\textbf {\bibinfo {volume} {99}},\ \bibinfo {pages} {1109}
  (\bibinfo {year} {1991})}\BibitemShut {NoStop}%
\bibitem [{\citenamefont {Scharfstein}\ and\ \citenamefont
  {Stein}(1990)}]{Scharfstein1990}%
  \BibitemOpen
  \bibfield  {author} {\bibinfo {author} {\bibfnamefont {D.~S.}\ \bibnamefont
  {Scharfstein}}\ and\ \bibinfo {author} {\bibfnamefont {J.~C.}\ \bibnamefont
  {Stein}},\ }\href@noop {} {\bibfield  {journal} {\bibinfo  {journal} {Am.
  Econ. Rev.},\ }\textbf {\bibinfo {volume} {80}},\ \bibinfo {pages} {465}
  (\bibinfo {year} {1990})}\BibitemShut {NoStop}%
\bibitem [{\citenamefont {Deneubourg}\ and\ \citenamefont
  {Goss}(1989)}]{Deneubourg1989}%
  \BibitemOpen
  \bibfield  {author} {\bibinfo {author} {\bibfnamefont {J.}~\bibnamefont
  {Deneubourg}}\ and\ \bibinfo {author} {\bibfnamefont {S.}~\bibnamefont
  {Goss}},\ }\Doi {10.1080/08927014.1989.9525500} {\bibfield  {journal}
  {\bibinfo  {journal} {Ethol. Ecol. Evol.},\ }\textbf {\bibinfo {volume}
  {1}},\ \bibinfo {pages} {295} (\bibinfo {year} {1989})}\BibitemShut {NoStop}%
\bibitem [{\citenamefont {van Kampen}(2007)}]{Kampen2007}%
  \BibitemOpen
  \bibfield  {author} {\bibinfo {author} {\bibfnamefont {N.~G.}\ \bibnamefont
  {van Kampen}},\ }\href@noop {} {\emph {\bibinfo {title} {{Stochastic
  Processes in Physics and Chemistry}}}},\ \bibinfo {edition} {3rd}\ ed.\
  (\bibinfo  {publisher} {Elsevier Science},\ \bibinfo {address} {Amsterdam},\
  \bibinfo {year} {2007})\BibitemShut {NoStop}%
\bibitem [{\citenamefont {Gillespie}(1977)}]{Gillespie1977}%
  \BibitemOpen
  \bibfield  {author} {\bibinfo {author} {\bibfnamefont {D.~T.}\ \bibnamefont
  {Gillespie}},\ }\href@noop {} {\bibfield  {journal} {\bibinfo  {journal} {J.
  Phys. Chem.},\ }\textbf {\bibinfo {volume} {81}},\ \bibinfo {pages} {2340}
  (\bibinfo {year} {1977})}\BibitemShut {NoStop}%
\bibitem [{\citenamefont {McKane}\ \emph {et~al.}(2013)\citenamefont {McKane},
  \citenamefont {Biancalani},\ and\ \citenamefont {Rogers}}]{McKane2013}%
  \BibitemOpen
  \bibfield  {author} {\bibinfo {author} {\bibfnamefont {A.~J.}\ \bibnamefont
  {McKane}}, \bibinfo {author} {\bibfnamefont {T.}~\bibnamefont {Biancalani}},
  \ and\ \bibinfo {author} {\bibfnamefont {T.}~\bibnamefont {Rogers}},\
  }\href@noop {} {\bibfield  {journal} {\bibinfo  {journal} {Bull. Math.
  Biol.}} (\bibinfo {year} {2013})},\ \bibinfo {note} {in press}\BibitemShut
  {NoStop}%
\bibitem [{\citenamefont {Ewens}(2004)}]{ewens2004mathematical}%
  \BibitemOpen
  \bibfield  {author} {\bibinfo {author} {\bibfnamefont {W.~J.}\ \bibnamefont
  {Ewens}},\ }\href@noop {} {\emph {\bibinfo {title} {Mathematical population
  genetics: I. Theoretical introduction}}},\ Vol.~\bibinfo {volume} {27}\
  (\bibinfo  {publisher} {Springer},\ \bibinfo {year} {2004})\BibitemShut
  {NoStop}%
\bibitem [{\citenamefont {Wallace}\ \emph {et~al.}(2012)\citenamefont
  {Wallace}, \citenamefont {Gillespie}, \citenamefont {Sanft},\ and\
  \citenamefont {Petzold}}]{wallace2012linear}%
  \BibitemOpen
  \bibfield  {author} {\bibinfo {author} {\bibfnamefont {E.~W.~J.}\
  \bibnamefont {Wallace}}, \bibinfo {author} {\bibfnamefont {D.~T.}\
  \bibnamefont {Gillespie}}, \bibinfo {author} {\bibfnamefont {K.~R.}\
  \bibnamefont {Sanft}}, \ and\ \bibinfo {author} {\bibfnamefont {L.~R.}\
  \bibnamefont {Petzold}},\ }\href@noop {} {\bibfield  {journal} {\bibinfo
  {journal} {IET Syst. Biol.},\ }\textbf {\bibinfo {volume} {6}},\ \bibinfo
  {pages} {102} (\bibinfo {year} {2012})}\BibitemShut {NoStop}%
\bibitem [{\citenamefont {Abramowitz}\ and\ \citenamefont
  {Stegun}(1965)}]{Abramowitz1965}%
  \BibitemOpen
  \bibfield  {author} {\bibinfo {author} {\bibfnamefont {M.}~\bibnamefont
  {Abramowitz}}\ and\ \bibinfo {author} {\bibfnamefont {I.~A.}\ \bibnamefont
  {Stegun}},\ }\href@noop {} {\emph {\bibinfo {title} {{Handbook of
  Mathematical Functions}}}}\ (\bibinfo  {publisher} {Dover},\ \bibinfo
  {address} {New York},\ \bibinfo {year} {1965})\BibitemShut {NoStop}%
\bibitem [{\citenamefont {Slater}(1966)}]{Slater1966}%
  \BibitemOpen
  \bibfield  {author} {\bibinfo {author} {\bibfnamefont {L.~J.}\ \bibnamefont
  {Slater}},\ }\href@noop {} {\emph {\bibinfo {title} {{Generalized
  Hypergeometric Functions}}}}\ (\bibinfo  {publisher} {Cambridge University
  Press},\ \bibinfo {address} {Cambridge},\ \bibinfo {year} {1966})\BibitemShut
  {NoStop}%
\end{thebibliography}%
\end{document}